\documentclass[aps,prab,twocolumn,superscriptaddress]{revtex4-1}
\usepackage{amssymb}
\usepackage{lipsum}
\usepackage{amssymb}
\usepackage{bbold}
\usepackage{amssymb}
\usepackage{graphicx}
\usepackage{graphics}
\usepackage{amsmath}
\usepackage{placeins}
\usepackage{hyperref}
\usepackage[dvipsnames]{xcolor}
\usepackage{wasysym}
\usepackage{soul}
\usepackage{float}
\usepackage{array}
\usepackage{tabulary}
\usepackage{caption}
\usepackage{fancyhdr}
\usepackage{wrapfig}

\usepackage[english,ngerman]{babel}

\usepackage[normalem]{ulem}
\usepackage{graphicx}
\usepackage{color,soul}
\usepackage{subfigure}
\usepackage{dcolumn}
\usepackage{bm}
\usepackage{hyperref}
\usepackage{url}
\usepackage{multirow}
\hypersetup{
    colorlinks=true,
    linkcolor=blue,
    filecolor=blue,      
    urlcolor=blue,
    citecolor=blue,
}

\usepackage[makeroom]{cancel}

\begin{document}
\selectlanguage{english}
\title{Dual-mode rf cavity: design, tuning and performance}

    \author{\firstname{Benjamin} \surname{Sims}}
    \email{simsben1@msu.edu}
    \affiliation{Department of Electrical and Computer Engineering, Michigan State University, MI 48824, USA}
    \affiliation{Department of Physics and Astronomy, Michigan State University, East Lansing, MI 48824, USA}
    \affiliation{Facility for Rare Isotope Beams, Michigan State University, East Lansing, MI 48824, USA}

    \author{\firstname{David} \surname{Sims}}
    %\email{simsdav4@msu.edu}
    \affiliation{Department of Electrical and Computer Engineering, Michigan State University, MI 48824, USA}
    \affiliation{Department of Physics and Astronomy, Michigan State University, East Lansing, MI 48824, USA}
    
    \author{\firstname{Sergey V.} \surname{Baryshev}}
    \email{serbar@msu.edu}
	\affiliation{Department of Electrical and Computer Engineering, Michigan State University, MI 48824, USA}
	\affiliation{Department of Chemical Engineering and Material Science, Michigan State University, MI 48824, USA}
    
    \author{\firstname{John W.} \surname{Lewellen}}
    %\email{jwlewellen@lanl.gov}
    \affiliation{Accelerator Operations and Technology Division, Los Alamos National Laboratory, NM 87545, USA}

\begin{abstract}
    We present the design and characterization of a dual-mode radiofrequency (rf) cavity, a novel electromagnetic structure with potential benefits such as compactness, efficiency, cost reduction and multifunctionality. The cavity was designed to balance the dual-mode structure considering several factors, such as mode frequencies, quality factor ($Q$-factor), and minimizing cross talk between couplers. We preformed various tests to verify that this cavity preformed as expected compared to simulated results. As exampled here, a combination of the the fundamental mode TM$_{010}$ and the TM$_{011}$ mode, tuned to a harmonic of the fundamental, was realized to linearize the off-crest electric field, thereby enabling concurrent bunching and acceleration of charged particle (e.g. electrons) beam in high power systems. The reduction in the number of cavities required to bunch and accelerate promises cost and space savings over conventional approaches. This research lays the foundation for further exploration of multi-mode cavity applications and optimization for specific use cases, with potential implications for a wide range of fields including quantum information platforms.

\end{abstract}

\maketitle

\section{Introduction}\label{intro}

When boundary conditions are applied to a wave, electromagnetic wave or a particle, travelling in free space, eigen modes or states form. Selection and manipulation over such modes or states are the "control knobs" in laser and information technologies, microelectronics, gaseous and solid state chemistry and beyond.
Cavities and waveguides are universally utilized tools for eigenmode control for information storage and transmission. Inherently, they can host/support many modes at once. This often is viewed negatively such as in a classical rectangular waveguide where 1/2 ratio of sidewall length is used to allow for one propagating TE$_{01}$ mode and avoiding crosstalk.

Multi-mode systems offer potential advantages of interest. In one example, when interacting with electrons in particular spin states, coexistence of many modes are extremely beneficial for information processing. It became a key subject in quantum computing and optics where superconducting radiofrequency (SRF) cavities, operated in the GHz range, are used to create and manipulate long lived quantum states  \cite{PhysRevLett.127.107701, Kollar_2015}. There is also a specific use of copper and niobium (normal and superconducting) rf cavities in fundamental research such as electron-proton colliders (particle physics), X-ray free electron lasers (basic energy sciences) or specialty electron linacs for space exploration \cite{10.3389/fspas.2020.00023}. The use of multimode cavities here were proposed to improve compactness/reduce cost or enhance luminosity/brilliance leading to much higher signal to noise ratio when detecting novel particles or exotic processes \cite{sicking_updated_2016, osti_1630267}. At the same time, multimode rf cavities (wherein mode frequencies are at simple integer or fractional ratios, and mode amplitude and phase are actively controlled) have been proposed \cite{dowell_two-frequency_2004, serafini_neutralization_nodate, lewellen_higher-order_2001, gong_design_2024} but generally have not been widely adopted.

Here, we present design of a dual-mode cavity that can simultaneously accelerate and shape electrons in linac systems, where amplitude and phase between the modes can be tuned with high precision. The demonstrated experimental feasibility of a dual-mode cavity has immediate translation to three-dimensional circuit QED architectures in which SRF cavities are being widely deployed as practical means for storing quantum states (transmon or photon) as they have millisecond to second scale coherence times \cite{PhysRevApplied.13.034032, wang_high-efficiency_2022}.

\section{Use of multi-mode Cavities in acceleration systems}\label{Multi-Mode}

One important system to consider is the rf photo-injector (or rf gun) which, e.g., serves to generate high quality electron beams for high brilliance hard X-ray production at FELs and time-resolved microscopy. The brilliance is highest when the greatest charge can be contained in the smallest phase space volume. Ultimately, brilliance would be limited by the Fermi exclusion principle and the Heisenberg's uncertainty principle. In reality, the beam quality from an rf gun is constrained by multiple factors, including those associated with the electron beam itself, i.e. non-linear space charge forces within the beam;  and factors associated with the design and performance of the gun's rf cavity, for instance the achievable electric field magnitude within the rf gun, and the sinusoidal time dependence of the fields that can distort the beam's phase space. To address challenges of this sort, specifically concerning improvement of the longitudinal phase space, a compelling idea of a double frequency rf cavity resonator was proposed \cite{dowell_two-frequency_2004, serafini_neutralization_nodate} where two modes, working cooperatively and tuned independently in amplitude and phase, couple to electrons in the beam to, e.g. linearize the longitudinal phase space. More generally multi-mode rf accelerator structures (including rf guns as well as structures intended to accelerate, but not generate, beams) have been explored in modeling and simulation, but very few have proceeded past that stage.

This paper presents the design and testing of a dual-mode TM$_{010}$/TM$_{011}$,  S/C-band (2.81/5.62 GHz) cavity. It was originally conceived with a particular focus of leveraging recent advances in solid state GaN-based amplifier technology for all-solid-state, compact accelerators. By using the second harmonic to linearize the time-dependence of the field experienced by a particle transiting the cavity, it can serve to both accelerate and chirp a beam for velocity-based bunch compression, while reducing or eliminating the intra-cavity beam energy droop typically associated with the operation of a standard "buncher" cavity.

\subsection{Conventional Bunching}
TM$_{010}$ cavities are commonly used for low-energy particle beam longitudinal compression. The phase of the TM$_{010}$ mode within a cavity is set so as to apply an energy chirp to a discrete bunch of particles (such as is generated by a photocathode and drive laser) transiting the cavity \cite{akre_commissioning_2008}, or to energy modulate a continuous beam (such as generated by a thermioic cathode). For a discrete bunch, the cavity phase is set such that the beam exits the cavity with no net energy gain or loss, but that the leading particles experience a net deceleration (energy loss) while the trailing particles experience a net acceleration (energy gain); the beam is thus given an energy chirp.  In the non- or quasi-relativistic regime, as the beam propagates the tail of the bunch catches up to the head of the bunch, increasing the peak current. The extent of such compression is dependent on the energy chirp provided via the buncher, the beam's average energy, and the distance propagated.  The ultimate compression achievable depends on space-charge effects, nonlinearities in the applied chirp, and magnitude of the applied chirp. While the net energy gain of the bunch exiting the cavity is, nominally, zero, as the beam particles transit the cavity they initially experience a net decelerating electric field and consequent reduction in energy. As the particles pass the midpoint of the cavity the magnitude of the electric field reduces to zero and reverses sign, reaccelerating the bunch. However, as space charge forces are enhanced at lower energies, the chirping process can result in beam emittance increases.  Further, the incoming energy of the beam limits the chirp that can be applied to the beam; attempting to apply too large of a chirp can physically reverse the beam's direction of motion.

To address this effect, bunchers are sometimes operated "off-zero", such that the beam experiences reduced deceleration through the first portion of the buncher cavity, and exits with a net higher energy than it entered.  Given the sinusoidal time dependence of the cavity field, however, the applied chirp becomes increasingly nonlinear the further "off-zero" the buncher cavity is operated.

\subsection{Multi-mode Bunching}
A buncher capable of supporting multiple TM mode, in contrast, breaks the tradeoff between deceleration and nonlinear chirping. Instead of synchronizing the particle bunches with the zero crossings of the rf field, the cavity modes are adjusted in phase and amplitude such that the incoming beam is accelerated, yet remains on a highly linear portion of the rf waveform. The result is a reduction (or elimination) of the beam deceleration while transiting the cavity; a net energy gain; and a highly linear chirp. Multi-mode bunching can thus mitigate the potential for emittance degradation due to deceleration; and it also allows for a stronger chirp to be applied to a lower-energy incoming beam, all else being equal, while maintaining a high degree of linearity in the applied chirp.  We note that while a net more-linear slope can be also achieved using two separate TM$_{010}$ cavities at two separate frequencies, while the net sum of the energy gain and chirp might be the same, two cavities apply the effects in sequence rather than concurrently.  Thus effects such as beam deceleration would still occur.

\section{Dual-mode Cavity Design}\label{Design}
The first step in designing the TM$_{010}$/TM$_{011}$ cavity was estimating the optimal length for both modes to coexist in a cylindrical pillbox cavity with a simple integer ratio between their mode frequencies, using the following equation \cite{lewellen_higher-order_2001}:

\begin{gather}
    f(L_{cav} , R_{cav} , p ,m ,n ) = \frac{c}{2 \pi}  \sqrt{\bigg(\frac{x_{p, m}}{R_{cav}}\bigg)^2 + \bigg(\frac{n \pi}{L_{cav}}\bigg)^2}   \label{eq:1} 
\end{gather}
The fundamental axial mode corresponds to the TM$_{010}$ mode with longitudinal and radial electric fields and a transverse (azimuthal) magnetic field. A similar field arrangement holds for the TM$_{011}$, except that this mode's on-axis electric field reverses direction at the midpoint of the cavity.  

The simple cylindrical cavity was then modeled in Superfish \cite{SUPERFISH} where an optimizer and sequencer where used to make fast and accurate changes to the cavity design to achieve both modes in the same geometry \cite{nasr_design_2018}. Since the frequency of the fundamental mode depends only on the cavity radius (e.g. $n$=0), while the TM$_{011}$ mode depends on both the cavity radius and length (e.g. $n$=1), in principle a cavity can be designed to support both modes at specific frequencies.
The cavity was designed to resonate at 2.81 GHz (TM$_{010}$) and 5.62 GHz (TM$_{011}$), i.e. an integer ratio of 2 between the fundamental and harmonic frequencies. The frequency choice was driven by two factors. First is the desire to build and demonstrate the new device in the technologically important S/C-band regime, where research on compact and high gradient cavities is presently very active \cite{simakov_update_2022, 10.1063/5.0132706}. Second is to benefit from the rapid advancements of high power solid state GaN amplifier technology to demonstrate beyond-klystron operation. Both then serve as a novel baseline for designing and building low cost, compact and high efficiency linac systems for industry, medicine and space exploration. 

The cavity shape was then modified from that of a simple cylindrical pillbox into a re-entrant design with the intent of increasing the cavity shunt impedance (for more effiicient rf power use), and to be consistent with current practice in high gradient, high efficiency cavity design \cite{puglisi_conventional_nodate}, again using an optimizer and sequencer. A study of the geometry was also preformed where the cavity profile was analyzed for the frequency response to wall displacements, to help inform frequency tuner placement; this topic is discussed further below. 

\begin{figure}
    \includegraphics[width=7cm]{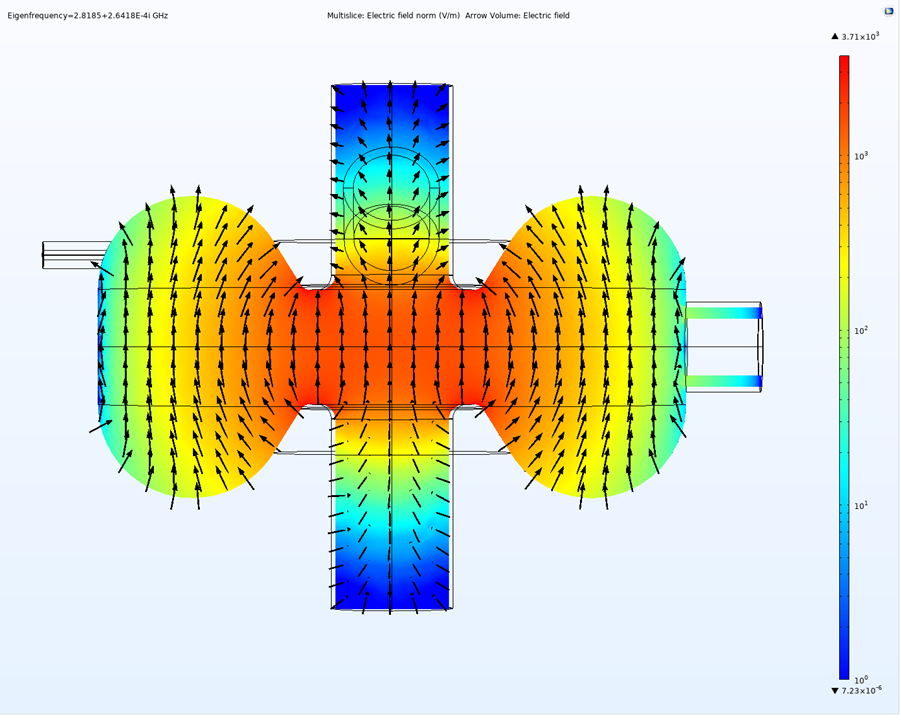}
	\caption{COMSOL model of fundamental mode electric field.}
	\label{COMSOL FUND EFIELD}
\end{figure}

\begin{figure}
	
 	\includegraphics[width=7cm]{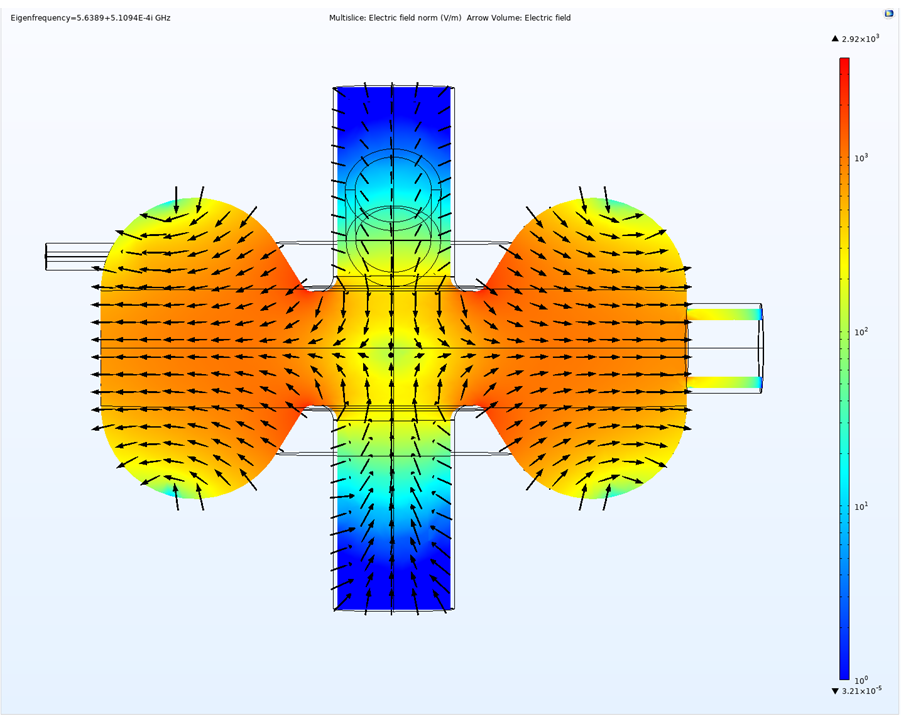}
	\caption{COMSOL model of TM$_{011}$ mode electric field.}
	\label{COMSOL HARM EFIELD}
\end{figure}

\begin{figure}
	\includegraphics[width=3cm]{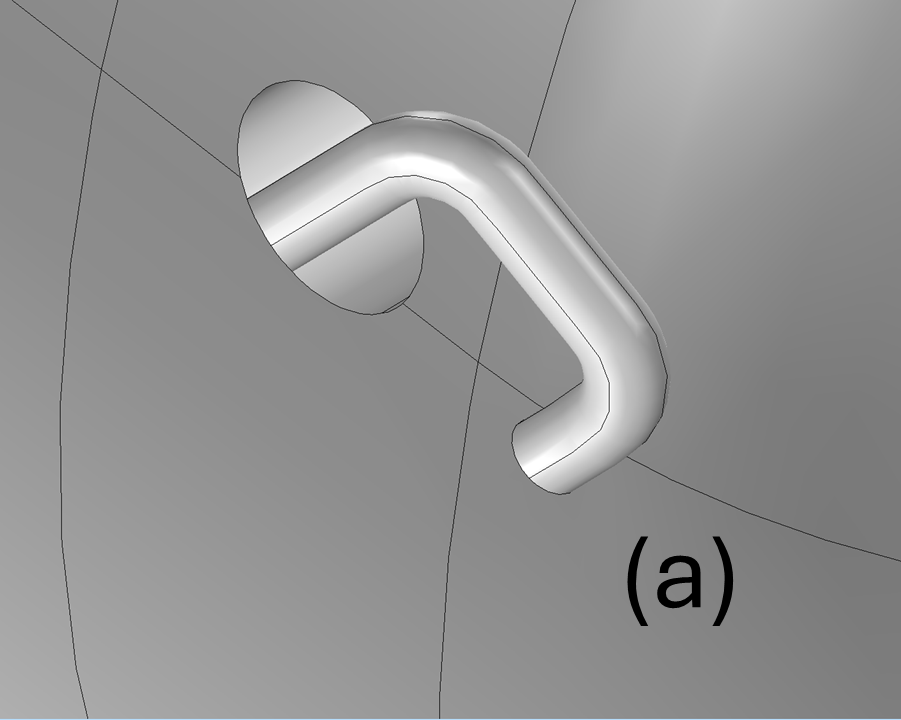}
    \includegraphics[width=3cm]{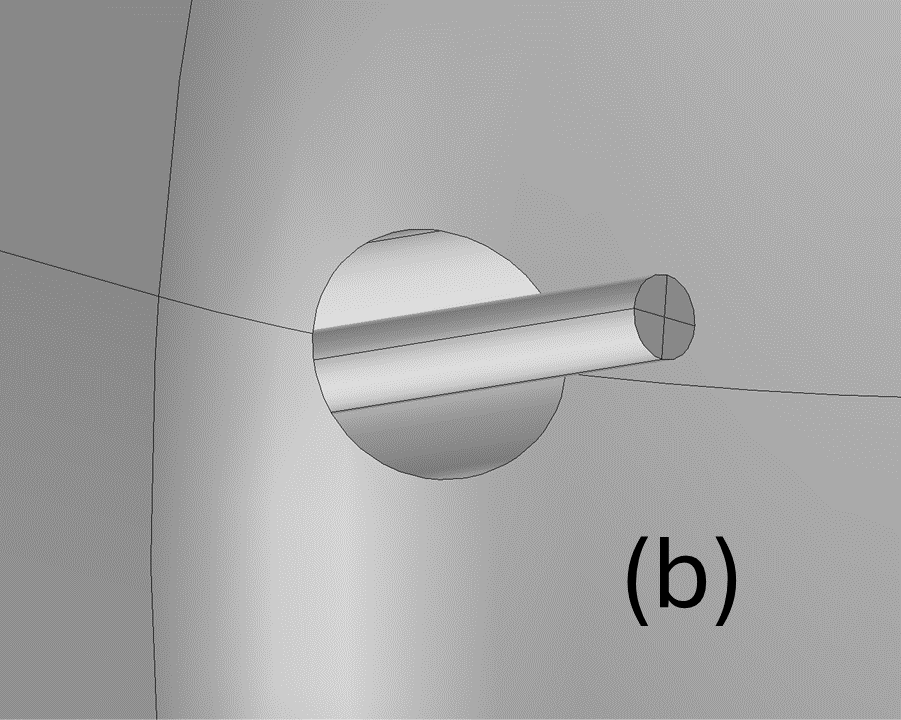}
	\caption{COMSOL models of coupling probes: (a) loop coupler and (b) electric field probe.}
	\label{COMSOL COUPLERs}
\end{figure}

\begin{figure}
	\includegraphics[width=5cm]{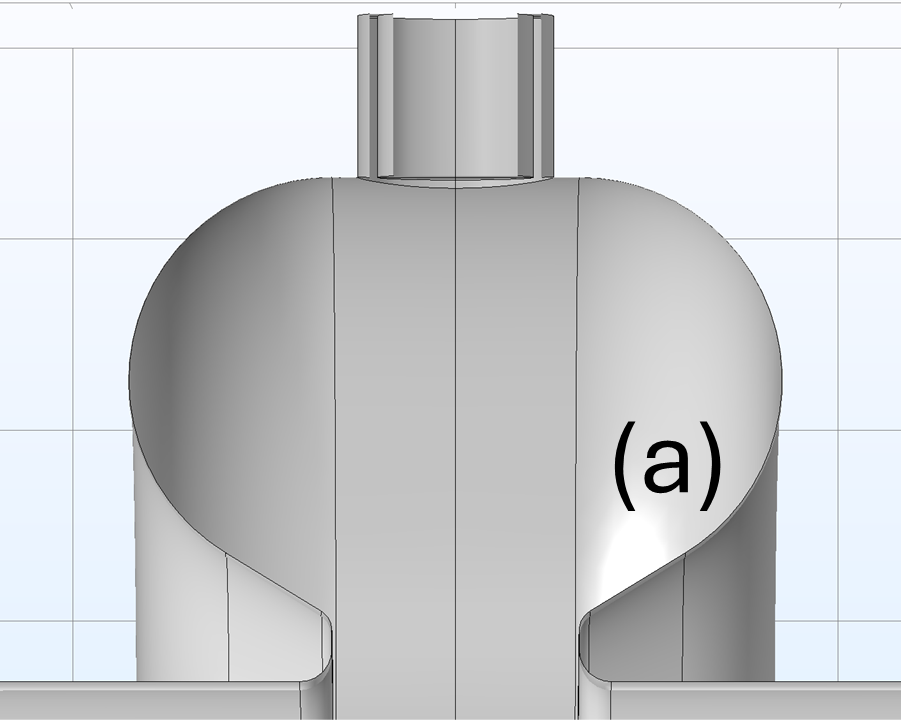}
	\includegraphics[width=5cm]{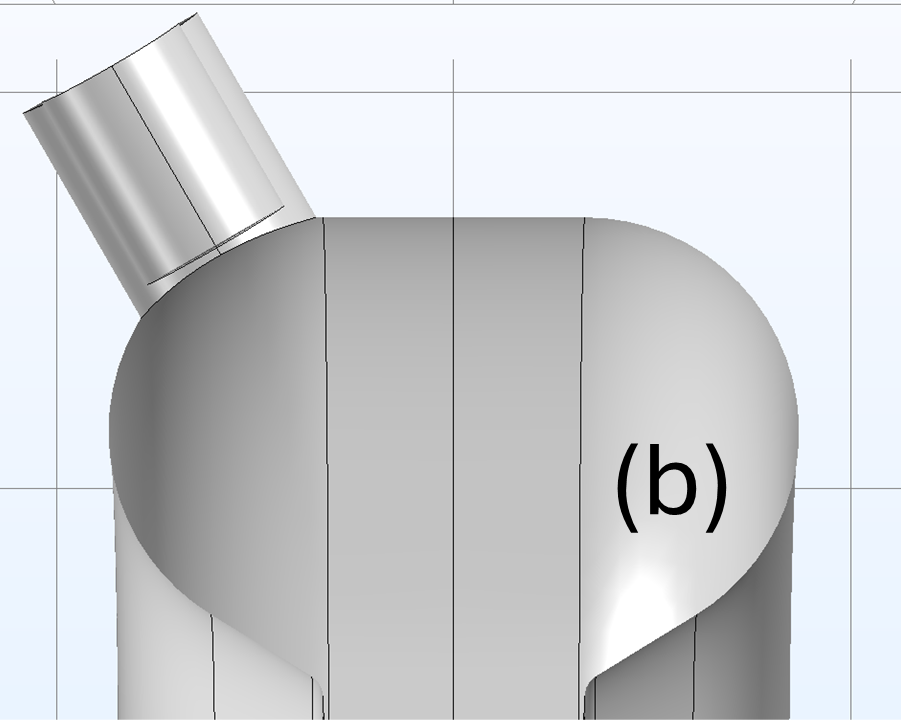}
	\caption{COMSOL models of tuners: (a) on-equator tuner and (b) off-equator tuner.}
	\label{COMSOL OFF AXIS TUNER}
\end{figure}

The reentrant cavity geometry was imported to COMSOL, a multiphyiscs code capable of performing 3D electromagnetic simulations (mode field distributions are shown in Figs.\ref{COMSOL FUND EFIELD}, \ref{COMSOL HARM EFIELD}). The model was retuned following the addition of tuner and RF power coupler ports. The design was such that each mode had its own independent coupler (Figs.\ref{COMSOL COUPLERs}). The tuner placements identified in the 2D Superfish simulations required very little modification, and finalized port arrangements are shown in Fig.\ref{COMSOL OFF AXIS TUNER}. Once tuner response, coupler cross talk, and eigenfrequency mode simulations proved satisfactory, the COMSOL 3D models were converted into full CAD drawings used for fabrication and brazing. 

\subsection{Couplers}

The cavity was designed based on the idea of using a separate coupler for each mode, which facilitates independent control of each mode's phase and amplitude and should simplify setting the coupling.
Each coupler was placed so that it primarily interacted with only one mode. The fundamental mode was excited via a loop \cite{puglisi_conventional_nodate, haebel_couplers_nodate} situated to couple strongly to the fundamental mode's magnetic field. Its location and orientation was optimized to exhibit minimal net coupling to the TM$_{011}$ mode.

Placement of the the TM$_{011}$ mode coupler was somewhat more straightforward to downselect. As the mode's (radial) electric field exhibits a relatively high amplitude at the cavity outer radius near the cavity equator, where the TM$_{010}$ mode field has a vanishing electric field along all directions in that region (see Fig. \ref{COMSOL HARM EFIELD}) an electric field probe \cite{puglisi_conventional_nodate, haebel_couplers_nodate} was an effective method to couple exclusively to the TM$_{011}$ mode. Its location was also optimized to account for the specifics of the cavity modes' field distributions, versus the ideal pillbox geometry. The specific method of multi-mode coupler design and placement is currently patent pending \cite {patent}.

Avoiding locations of maximum field produced by each mode, and instead balancing all field at a given location, helped minimize cross-talk between the couplers. This is important in a two coupler system as strong cross talk can be detrimental to system performance.  Couplers with significant crosstalk would both lower the overall cavity efficiency (e.g. in terms of rf power required to obtain a given accelerating gradient) due to undesired power out coupling, but also necessitate the rf network having a highly robust and effective, harmonic rf isolation system between the cavity and rf power sources.

\subsection{Tuners}
Maintaining synchronization between the modes (e.g. integer or simple-fraction frequency ratios) is critical for maintaining coherence between them and beams transiting the cavity. For our cavity, ideally, the modes should be integer multiples of one another, specifically the frequency of the TM$_{010}$ should be exactly 1/2 of the frequency of the TM$_{011}$ mode.  (Small differences in frequency ratios can in principle be addressed by, for instance, increasing the input power and operating slightly off the natural resonance; but such approaches depend on the magnitude of the ratio error and the cavity bandwidths.  It is thus generally preferred to have the frequency ratio as close to ideal as possible from the start.) To further optimize the performance of the cavity, tuners have been implemented: an on-equator tuner and an off-equator tuner. The tuners are plungers \cite{longuevergne_innovative_2014,veshcherevich_rf_1995} and together enable independent adjustments to the resonant frequencies of the two modes within the cavity. Inserting or retracting the tuners alters the cavity volume \cite{veshcherevich_rf_1995} and thus the frequencies of the modes.
The on-equator tuner, when inserted, effects a positive change in frequency of TM$_{011}$ with a concurrent negative change in frequency of TM$_{010}$. The off-equator tuner, on the other hand, increases the resonant frequency of both modes when inserted. The relative insertion depth of the two tuners can, within limits, be set to attain the desired integer ratio of 2 between the TM$_{010}$ and TM$_{011}$ resonant frequencies, as well as set an absolute frequency for one of the modes.

\subsection{Bead Pulls of the On-Axis Electric Field}

\begin{figure}
	\includegraphics[width=7cm]{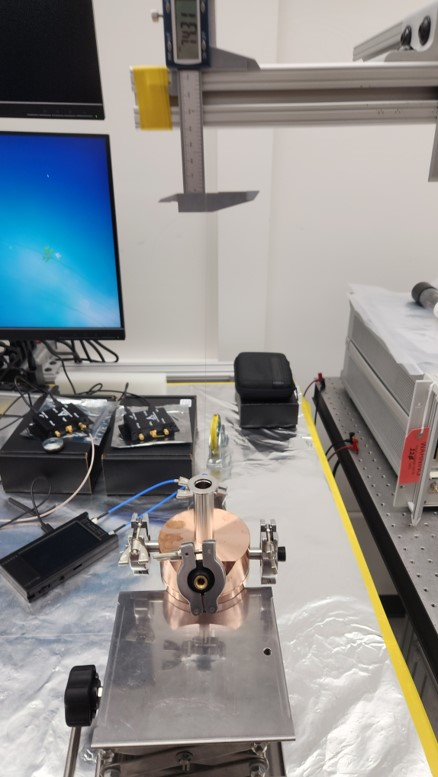}
	\caption{Experimental bead pull setup}
	\label{BEADPULL}
\end{figure}

\begin{figure}
	\includegraphics[width=7cm]{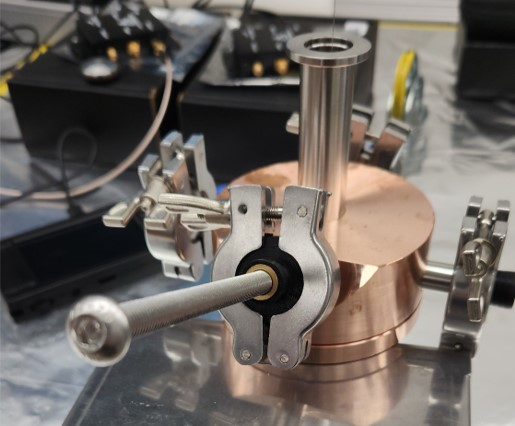}
	\caption{Experimental tuner setup }
	\label{TUNER}
\end{figure}

\begin{figure}
	\includegraphics[width=7cm]{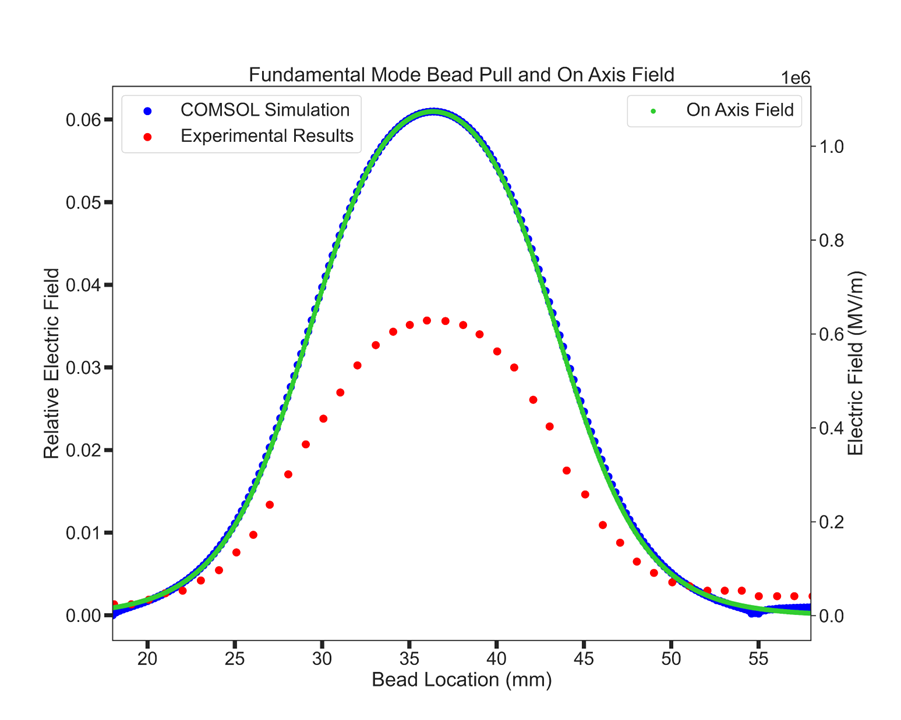}
	\caption{Fundamental mode bead pull results (left axis) compared to on-axis field (normalized at 1 MV/m) taken from Superfish (right axis).}
	\label{FUND}
\end{figure}

\begin{figure}
	\includegraphics[width=7cm]{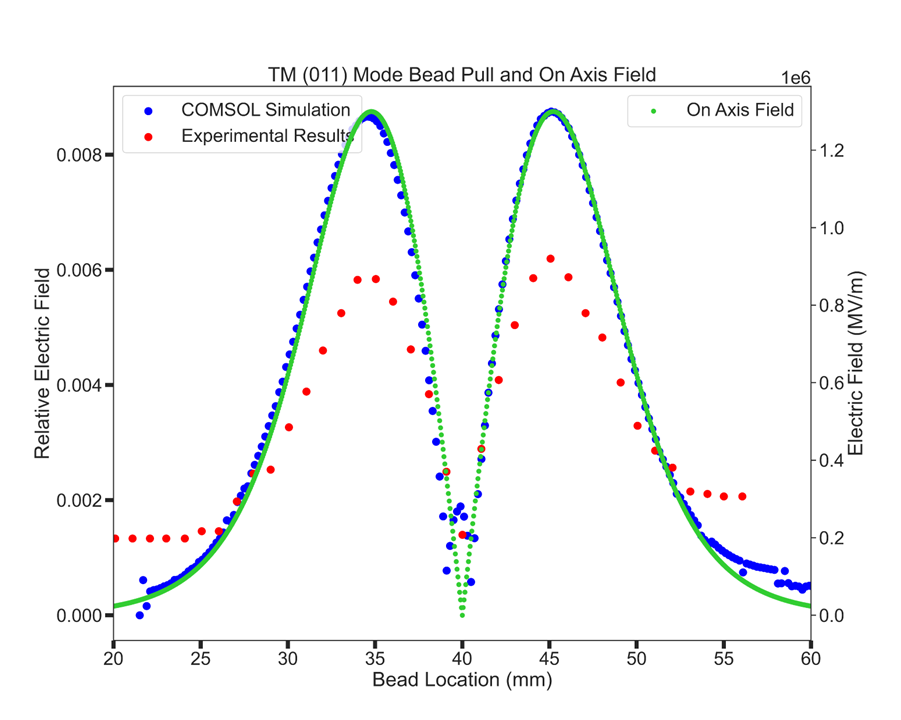}
	\caption{TM$_{011}$ mode bead pull results (left axis) compared to on-axis field (normalized at 1 MV/m)taken from Superfish (right axis).}
	\label{HARM}
\end{figure}

Perturbing the electromagnetic fields in a cavity in a small volume, using for instance a small metal bead pulled through the cavity, will change the resonant frequency of a mode based upon the magnitude of the perturbation \cite{RevModPhys.18.441, palmer_next_2005}.  This allows the mode's spatial field pattern to be mapped, in the limit of small perturbations \cite{meekes_characterizing_nodate, wegner_bead-pull_2014, dowell_chapter_nodate,callebaut_field_1960, pozar_microwave_2012,corlett_measurement_1993}. When mapping the on-axis fields of the mode, where the magnitude of the magnetic field is negligible, the relative magnitude of the unperturbed electric field $E_p$ within the perturbed volume is related to the mode frequency change simply by 

\begin{gather}
    E_p = \sqrt{\frac{\omega - \omega_0}{\omega_0}} \label{eq:2} 
\end{gather}
This method allows for mapping the on-axis fields of both modes to verify the field pattern predicted by COMSOL, and also to verify the mode identification performed via network analyzer.  A bead consisting of a 16 gauge (1.3 mm diameter) 3 mm long section of copper wire attached to a dielectric string was pulled through the cavity in 1 mm increments along the longitudinal axis, and the cavity mode frequency measured as a function of the bead's position. These measurements were performed for both modes. A simulated bead pull was also performed using COMSOL where a similarly sized cylinder was used as a stand-in for the wire section. This allows for a direct comparison between the experimental bead pull and a simulated bead pull. 
Comparing the results, it is seen that there are certain features of the field that are not captured by simple 2D simulations, nor by simply comparing the on-axis field from an eigenmode solver to the beadpull measurement.  For instance, the differences in field magnitudes at the cavity/beamtube boundary is likely due to asymmetries introduced from adding the tuners and rf power couplers. The harmonic bead pull data are also noisy near the zero crossing; this is expected from the nature of the perturbation-based measurement.
    The fields near the cavity boundaries are particularly difficult to measure due to the low field amplitude; this can be seen from superimposing the simulation and the experimental data. Overall, there is an excellent agreement between the simulated and experimental $E_p$ results as shown in  Fig.~\ref{FUND}~and~Fig.~\ref{HARM}.  A variation on this method was used previously with a larger bead and multiple tests which also saw similar results \cite{sims:ipac2024-mocd2}.

\subsection{Measured Q-factor and Mode Frequencies}

The quality factor was measured using a vector network analyzer (VNA) within the standard $S_{21}$ approach \cite{kajfez_reflection-type_2000, ginzton_microwave_1957,dowell_chapter_nodate, haebel_couplers_nodate}, where $Q_L$ (loaded quality factor) was determined by measuring the resonant frequency $\omega_0$ and determining the bandwidth $\Delta \omega$ (using two --3~dB points off of the $S_{21}$ peak):
\begin{gather}
    Q_L= \frac{\omega_0}{\Delta\omega}  \label{eq:3}
\end{gather}

An electric field probe placed in the beam port along the beam axis was used as the pickup probe for all $Q$-factor measurement. Once the desired probe was inserted into its respective port and coupled strongly to the cavity, the $Q$-factor was measured based on the $S_{21}$ response. The pickup probe was shortened and the $Q$-factor was again measured. This was repeated to achieve minimal coupling with the pickup probe to ensure that the coupling parameter $\beta$ was equal to 1 for the input coupler. Multiple measurements, including removing and replacing the couplers, yielded $Q_0$=8--9k, where  $Q_0 = Q_L\cdot(1 + \beta)$ \cite{gregory_q-factor_2022}.
These results were reasonably close to the computed values of $\sim$11k from COMSOL and speak to the high fabrication quality of the cavity. The resonant frequencies for both modes were within $~10$ MHz from their design values.

\begin{table}[!]
\centering
\caption{Simulation (sim) and experimental (exp) $Q$-factors and resonant frequencies (with tuners removed).}
\begin{tabular}{ l | c c c  | c  c  c}

Parameter &       & TM$_{010}$ & & &TM$_{011}$ \\ 
\hline
\hline
& exp & sim  && exp   &  sim   \\
\hline
$Q$-factor & 8407 & 10661 & &  9362  &  11299  \\
\hline
$f_0$ (GHz) & 2.8166 &  2.8186& & 5.6267  &  5.6388 \\

\end{tabular}
\label{T:1}
\end{table}

\begin{figure}
	\includegraphics[width=7cm]{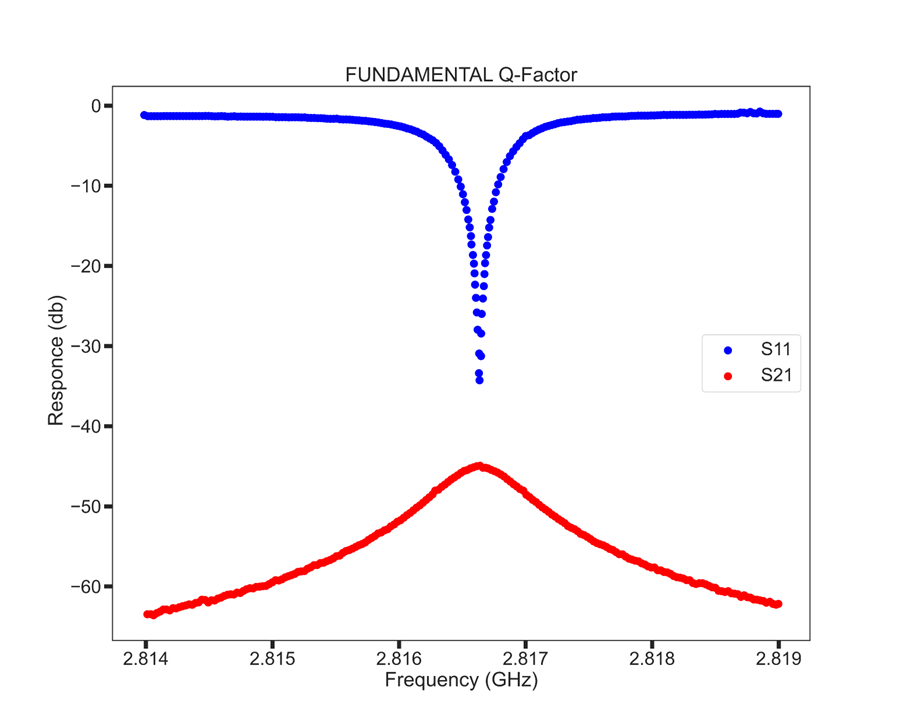}
	\caption{Experimental $S_{11}$ and $S_{21}$ used to measure fundamental mode $Q$-factor.}
	\label{FUND_Q}
\end{figure}

\begin{figure}
	\includegraphics[width=7cm]{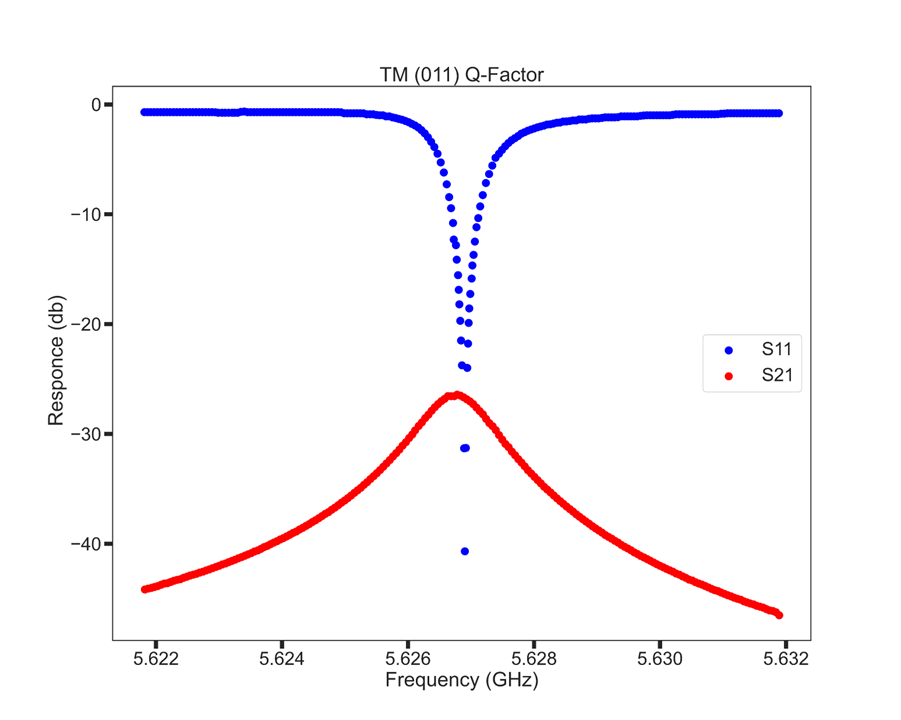}
	\caption{Experimental $S_{11}$ and $S_{21}$ used to measure TM$_{011}$ mode $Q$-factor.}
	\label{HARM_Q}
\end{figure}

\subsection{Mode Cross Talk Testing}

\begin{figure}
	\includegraphics[width=7cm]{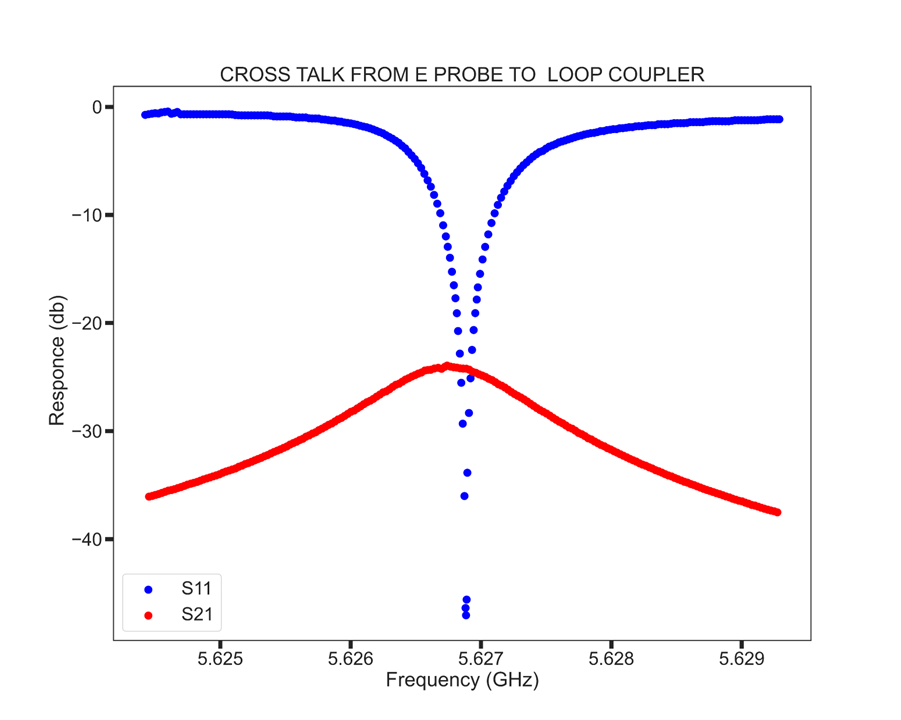}
	\caption{Experimental $S_{11}$ and $S_{21}$ used to measure cross talk from electric field probe to loop coupler.}
	\label{E_2_LOOP}
\end{figure}

\begin{figure}
	\includegraphics[width=7cm]{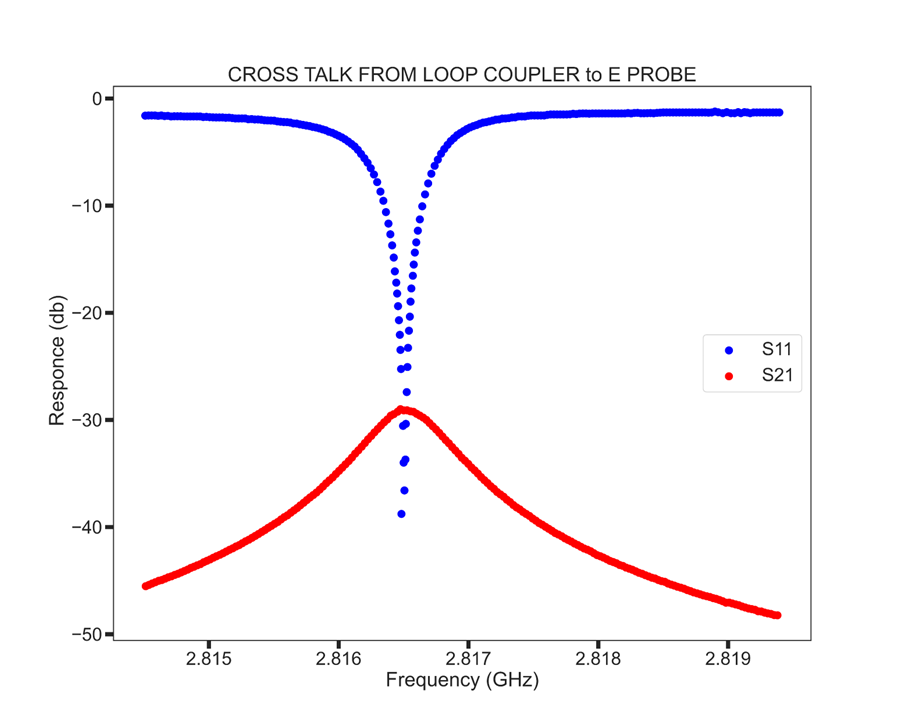}
	\caption{Experimental $S_{11}$ and $S_{21}$ used to measure cross talk from loop coupler to electric field probe.}
	\label{LOOP_2_E}
\end{figure}

As noted previously, in our design for operating two independent couplers in a single cavity, each is intended to have $\beta=1$ coupling to the desired mode and $\beta\approx$~0 for the other driven mode. To make a practically usable device it must be ensured that there is minimal power transmitted between the couplers. Power couplers which are strongly interacting with both their own and other powered modes (cross talk in other words) would result in potentially significant fractions of power being uselessly transmitted between the couplers with net diminished power coupled to the cavity volume. As this was an anticipated issue, coupler locations and geometries were modelled and optimized so as to minimize the cross talk. The fabricated multimode cavity, was used to characterize the crosstalk. The results shown in Figs. \ref{E_2_LOOP} and \ref{LOOP_2_E} highlight the low transmitted power between the couplers. When calculated from the log-scale, approximately 2\%  of power at f=2.81652 GHz was transmitted from the loop coupler to the electric field probe and 0.3\% of power at f=5.62667 GHz was transmitted from the electric field probe to the loop coupler. The crosstalk could likely be reduced further with additional optimization, however, these levels are practically very adequate to our present requirements. In practice, the adjustment process is somewhat tedious, and future studies should incorporate efforts to simplify the design and streamline the tuning process. When the loop coupler's coupling to the TM$_{011}$ mode was explored, as opposed to the designed-for fundamental mode, it was found, as expected, that the coupling was dependent upon the orientation of the loop coupler. The electric field probe, intended to couple to the TM$_{011}$ mode exhibited almost no coupling to the TM$_{010}$ mode, due to the minimal electric field magnitude that mode exhibits at the cavity equator, also as expected. These observations are consistent with the crosstalk measurements.

\subsection{Tuner Response}

\begin{figure}
	\includegraphics[width=7cm]{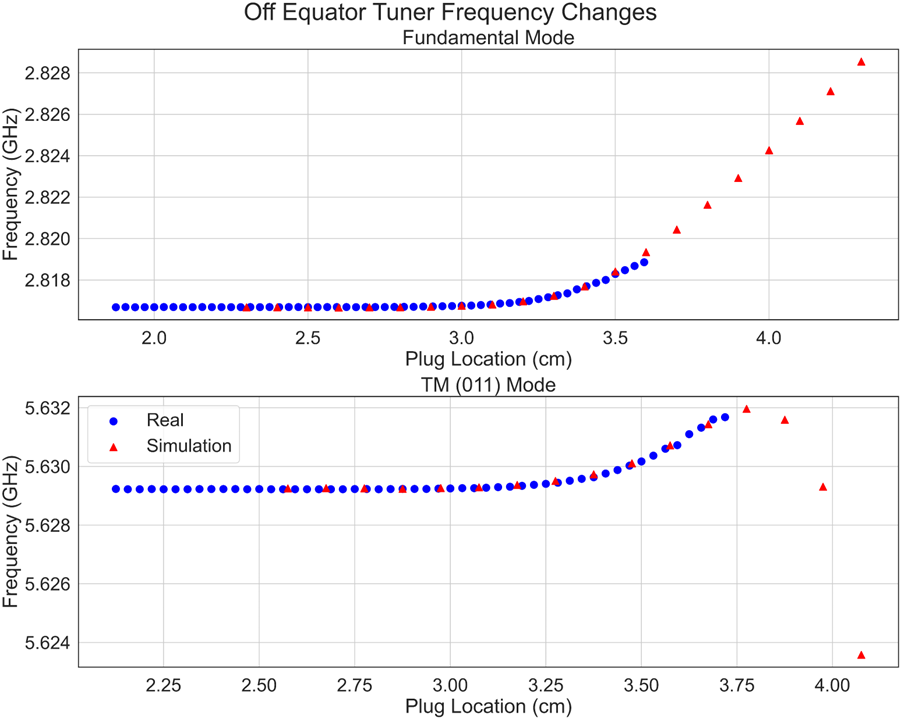}
	\caption{Frequency response from the off-equator tuner.}
	\label{Off Axis}
\end{figure}

\begin{figure}
	\includegraphics[width=7cm]{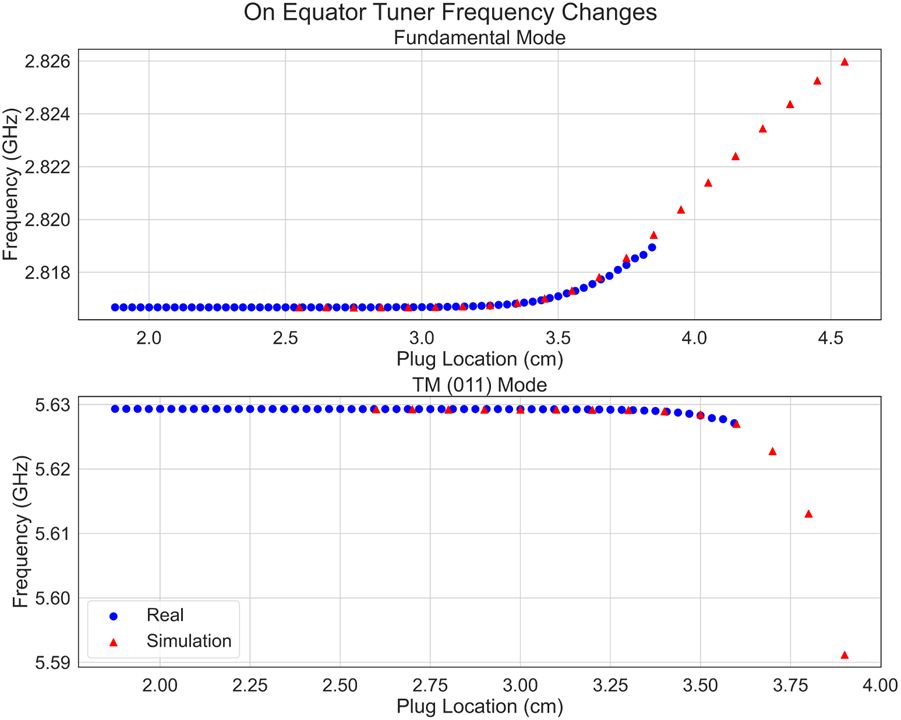}
	\caption{Frequency response from the on-equator tuner.}
	\label{On Axis}
\end{figure}

The resonant frequencies of the TM$_{010}$ and TM$_{011}$ cavity  modes were recorded as the tuners were moved in $\sim$0.3 mm steps. The tuner plugs were mounted and zeroed at the flange connection for each port prior to moving them. These tests were intended to determine the tuning range accessible by these tuners: with either $(i)$ the tuner being pulled out of the cavity so far it no longer had any effect on the resonant frequency, or $(ii)$ the tuner inserted far enough into the cavity to detune a mode to the point of significantly altering the field pattern and/or impacting its coupling. The results of the tuner tests are summarized in Figs.~\ref{Off Axis}~and~\ref{On Axis}, along with simulated runs from COMSOL. The as-designed and actual operating results are in excellent agreement.

\begin{figure}
	\includegraphics[width=7cm]{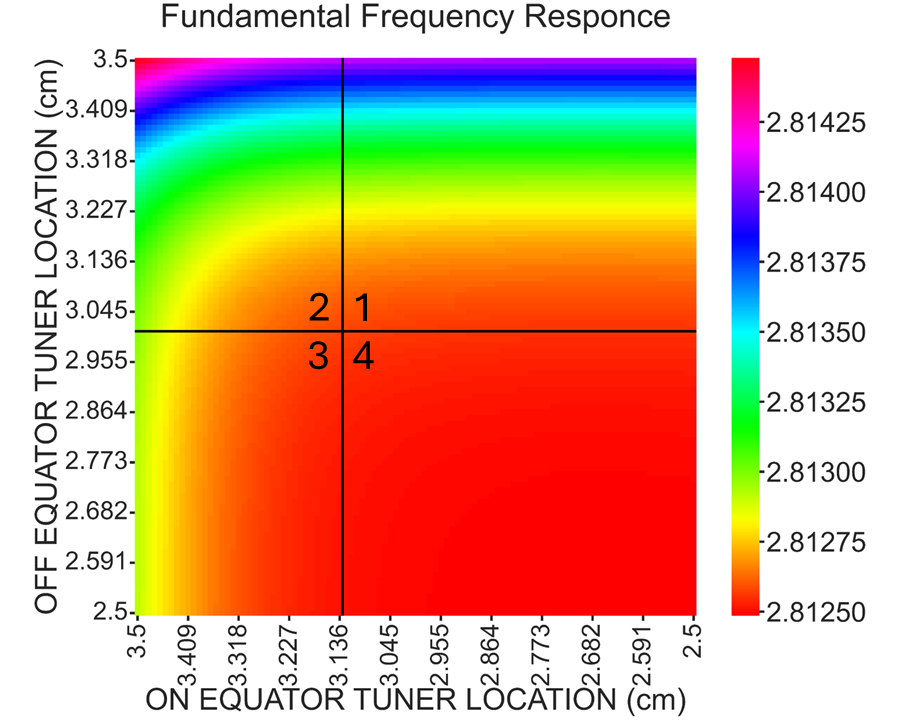}
	\caption{Heat map of the fundamental frequency based off of tuner locations. }
	\label{Fund heatmap}
\end{figure}

Using this data, we can better visualize tuner performance by creating heat maps which illustrate the individual tuners' effects as well as their combined effects. Starting with the TM$_{010}$ frequency response in Fig.\ref{Fund heatmap}, the black vertical and horizontal lines identify the tuner "zero position" -- approximately where the tuner is level with the cavity wall, and further insertion begins to significantly perturb the mode volumes. These lines let us logically divide the heat map into four quadrants. Quadrant \#4 is where both tuners are retracted past the cavity wall, and there is negligible frequency change from either tuner's motion. Quadrant \#1 captures the effect of inserting the off-equator tuner only which mirrors the expected frequency change seen in Fig.~\ref{Off Axis}:  the frequency increases as the tuner insertion depth increases. Quadrant \#3 captures the effect of inserting only the on-equator tuner and the result mirrors the effect seen in Fig.~\ref{On Axis} where the frequency increases as the tuner is inserted into the cavity. Finally, quadrant \#2 captures the effect of both tuners being inserted concurrently. This performance is as expected given the tuners' independent responses.

\begin{figure}
	\includegraphics[width=7cm]{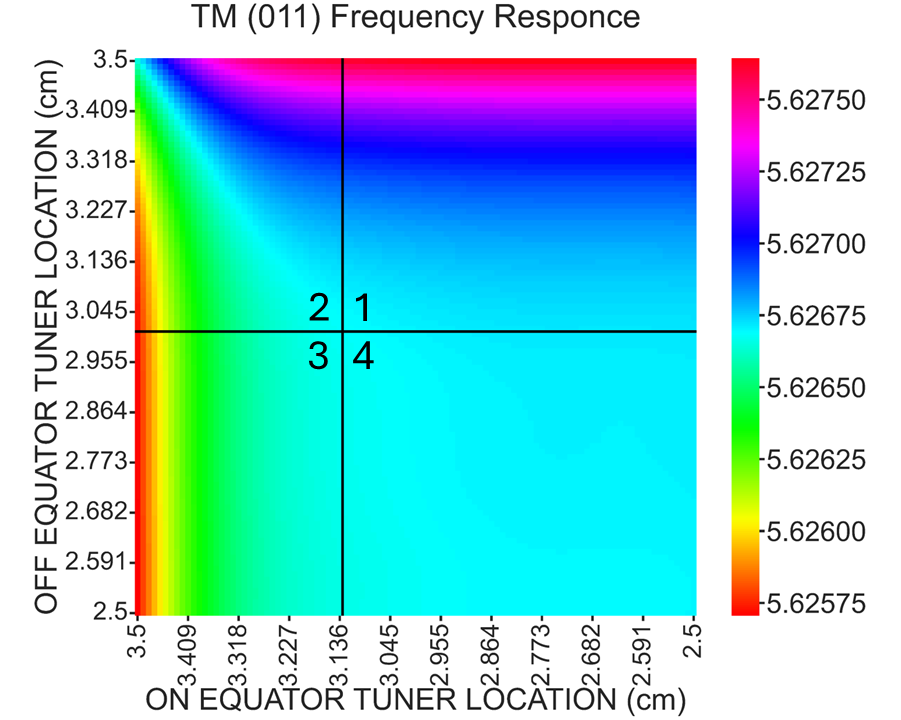}
	\caption{Heat map of the harmonic frequency based off of tuner locations.}
	\label{Harm heatmap}
\end{figure}

The same mapping and visualization approach was used for characterizing the TM$_{011}$ mode frequency response. The black horizontal and vertical lines label zero tuner locations in Fig. \ref{Harm heatmap}, however there is greater frequency response within quadrant \#4 as compared to that seen in the fundamental mode, see Fig. \ref{Fund heatmap}. This is expected due to the higher frequency mode's evanescent field penetrating deeper into the tuner ports than the fundamental's. As the frequency response for TM$_{011}$ mode is not in the same direction, quadrant \#1 and quadrant \#3 now show opposite frequency changes, as compared to the results in Figs.~\ref{On Axis}~and~\ref{Off Axis}. Quadrant \#2 captures the combined effects of two tuners being inserted into the cavity.

Tuners can only modify the frequency in limited amounts before their effects are detrimental to the cavity performance, e.g. excessive losses on the tuner, coupling perturbation, etc. Thus designing a multi mode cavity to have integer multiple mode frequencies from the start is critical to the final performance; ideally following fabrication, the cavity would have both a fundamental mode frequency and mode frequency ratios exactly at the design values with the tuners flush with the cavity walls. In practice, some manufacturing deviations are to be expected, and therefore the ability to perform frequency adjustment is necessary. 

\begin{figure}
	\includegraphics[width=7cm]{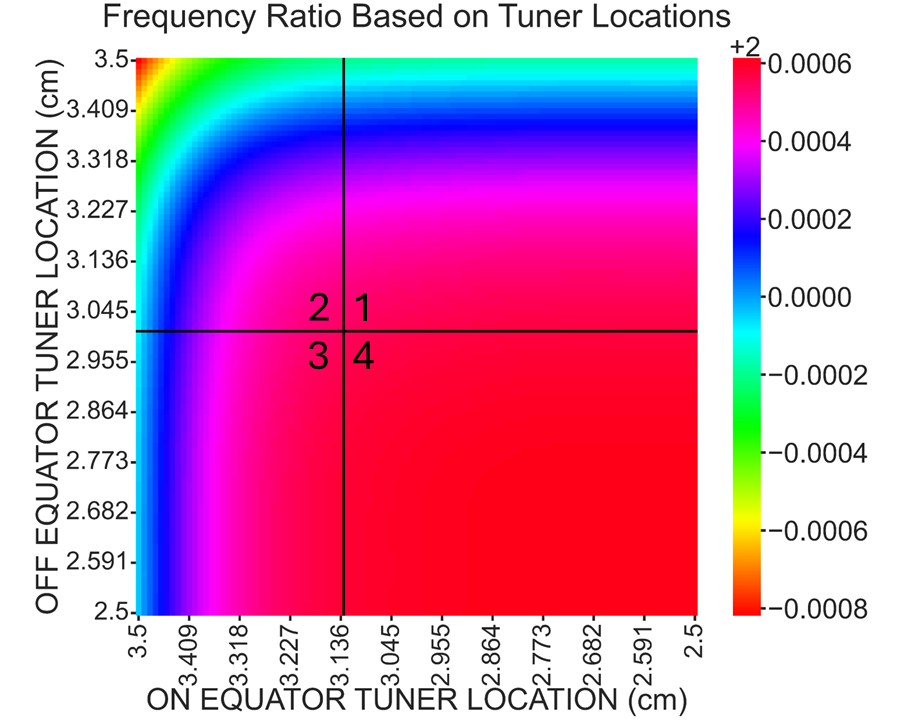}
	\caption{Frequency ratio between TM$_{011}$ and fundamental modes based off of tuner locations. }
	\label{INT MULT}
 
\end{figure}

Taking the ratio of the harmonic map and the fundamental map, a new map of the frequency ratio can be created, as seen in Fig.\ref{INT MULT}.
\begin{gather}
    {Frequency} {Ratio}= \frac{TM_{011}}{TM_{010}}  \label{freqratio}
\end{gather}
It shows the frequency ratio and gives the locations of the tuners which can then be used to tune this cavity to the desired $n=2$ integer. As observed from the tuner response results, the designed and fabricated system can increase the fundamental mode's frequency significantly. Ignoring, for the moment the change to TM$_{011}$, the TM$_{010}$ frequency can be increased to the point when it is at an integer multiple of a fixed TM$_{011}$ frequency. Thus, a decrease in the frequency ratio of the system is easily achievable, i.e. $n=2.1$ being lowered to $n=2$ . This is not true for the TM$_{011}$ frequency. If now change of TM$_{010}$ is ignored, it can be seen that due to the opposing response for each tuner it is easiest to maintain the frequency of TM$_{011}$ instead of making it strongly increased or decreased. These observations lead to the conclusion that deviation from the target TM$_{011}$ frequency is much harder to compensate for, than deviation of TM$_{010}$ frequency. Combining the responses at both frequencies together, it can be found that this system has the ability to decrease the frequency ratio between the modes by maintaining the TM$_{011}$ frequency and increasing the TM$_{010}$ frequency. To summarize, the cavity can be more easily tuned if f$_{011}$/f$_{010}$ $>$ 2 than  if  f$_{011}$/f$_{010}$ $<$ 2. 
While an analysis of cavity response and potential fabrication errors is normally a part of any cavity fabrication process, the above discussion illustrates additional considerations that should be taken into account when designing multifrequency multimode structures.

\section{Conclusions and Outlook}\label{Consclusion}

A dual-mode cavity was successfully designed, fabricated and tested. Its measured performance is in good agreement with  simulation results. This cavity amalgamates functionality typically distributed across several cavities (acceleration, bunching/chirping and linearization) into a single resonant structure, conserving valuable space in a beamline and providing potential performance improvements. Its relative simplicity in implementation and inherent ability to host other, even higher-order, modes with small deviations in design broadens applicability of the presented cavity across various applications. For example, higher-order modes can be used to decrease energy spread induced on a pre-bunched beam following bunch compression, or storing or reading out information in quantum information SRF platforms.

The design approach offers the benefits of compactness, cost and improved performance/efficiency within a minimal footprint, These are important considerations for applications where space and mass are at a premium, such as accelerators intended for spaceflight; but also can be important in laboratory settings, medical devices, or industrial environments. A compact beamline not only facilitates easier integration into existing infrastructure but also enhances portability, if needed, providing greater flexibility in deployment.

\section{Acknowledgments}\label{Acknow}
The work by Benjamin Sims was supported by the U.S. Department of Energy Office of Science, High Energy Physics under Cooperative Agreement Award No. DE-SC0018362. The work by Sergey Baryshev was supported by the U.S. Department of Energy, Office of Science, Office of High Energy Physics under Award No. DE-SC0020429. 

\bibliography{references}

\appendix
\addcontentsline{toc}{section}{Appendices}
\renewcommand{\thesubsection}{\Alph{subsection}}

\end{document}